
\magnification=1200
\baselineskip=18pt
\normallineskip=8pt
\overfullrule=0pt
\vsize=23 true cm
\hsize=15 true cm
\voffset=-1.0 true cm
\font\bigfont=cmr10 scaled\magstep1
\font\ninerm=cmr9
\footline={\hss\tenrm\folio\hss}
\pageno=1
\newcount\fignumber
\fignumber=0
\def\fig#1#2{\advance\fignumber by1
 \midinsert \vskip#1truecm \hsize14truecm
 \baselineskip=15pt \noindent
 {\ninerm {\bf Figure \the\fignumber} #2}\endinsert}

\def\ref#1{$^{[#1]}$}
\def\sqr#1#2{{\vcenter{\vbox{\hrule height.#2pt
   \hbox{\vrule width.#2pt height#1pt \kern#1pt
   \vrule width.#2pt}\hrule height.#2pt}}}}

\noindent{preprint HLRZ 24/93}
\bigskip
\bigskip
\centerline{\bigfont A SIMPLE GEOMETRICAL MODEL FOR SOLID FRICTION}
\bigskip
\bigskip
\bigskip
\bigskip
\smallskip
\centerline{{\bf T. P\"oschel$^{1,2}$ and H.J. Herrmann$^{1}$}
{\footnote{$^{\dagger}$}{on leave from C.N.R.S., France}}}
\medskip
\centerline{$^{1}$ HLRZ, KFA J\" ulich, Postfach 1913}
\centerline{5170 J\" ulich, Germany}
\smallskip
\centerline{$^{2}$ Physics Department, Humboldt University}
\centerline{ Invalidenstr. 42, 1040 Berlin, Germany}
\bigskip
\bigskip
\bigskip
\noindent{\bf Abstract} \par
\smallskip
We present a simple model for the friction of two solid bodies moving
against each other. In a self consistent way we can obtain
the dependence of the macroscopic friction force as a function
of the driving velocity, the normal force and the ruggedness of
the surfaces in contact. Our results are discussed in the context
of friction laws used in earthquake models.
\bigskip
\bigskip
\leftline{PACS numbers: 46.30.Pa 81.40.Pa 05.60.+w}

\vfill\eject
The friction of two solid bodies against each other has been studied
for many centuries. Coulomb\ref {1} formulated the law that the friction
force $F_{fr}$ slowing down the relative motion between the two bodies
is proportional to the normal force $F_n$ and the proportionality
constant is called the friction coefficient $\mu$.
In particular, within this law, the friction force is
supposed to be independent
on the area of contact and on the relative velocity $v$ with which
the two bodies move against each other. Experimentally it is, however,
well known\ref{2,3} that $F_{fr}$ does depend on $v$.

A more detailed knowledge of the dependence of $\mu$
on the shearing velocity $v$ is very important in earthquake
models. Using a function of the form
$$\mu(v) = \beta/(v+v_0)\eqno(1a)$$
it has been shown\ref{4} that the classical block-spring
model of Burridge and Knopoff\ref{5} is able to reproduce
realistic events including
the Gutenberg-Richter law for the distribution of earthquake
sizes. A friction law of the form
$$\mu(v) = \mu_0 + \beta \cdot ln ({v \over v_0})\eqno(1b)$$
has been introduced\ref{6} and used to describe slip motion
of faults\ref{7}. The question
arises: Is it possible to obtain the friction laws of eqs.~1
also from a microscopic model?

To that purpose one has to know the microscopic origin of friction.
Already in 1699 Amontons and de la Hire\ref{3} proposed
that the main contribution to solid friction
just comes from the geometrical ruggedness of the two
surfaces in contact. Since the two surfaces are not perfectly flat
they cannot move at the same time straight and maintain perfect
contact. The surfaces must separate a distance at least as far apart
as their highest asperities and they will jump up and down.
This vertical jumping will be attenuated by the elasticity of the
bulk but on a local level the surfaces will move apart when
asperities hinder each other and come close when their indentations
happen to fit better into each other.

Of course if one goes into detail many more physical effects
come into play\ref{2,3}. Enormous forces act at the contact of the tips
of asperities so that these tips are plastically deformed
and can even fragment off (wear). The fragments and eventual
intermediate fluids can lubrify the contacts, i.e. form
layers that smoothen the surfaces. The fragments can even exert the
role of bearings\ref{8}. Also on the molecular level
adhesion or electrostatic forces can play an important role.

Unfortunately, however, even the simplest problem, namely the purely
geometrical hindrance of the two surfaces sliding against each other
is already very difficult to deal with and from the theoretical
point of view very little is yet known. It is the purpose of this work
to shed more light on its understanding. In fact,
this simple situation can be realized by considering infinitely
strong and stiff elastic bodies in vacuum.
Experimentally a good material for that purpose would be rocks.

In the last years the roughness of solid surfaces has attracted
major interest mainly due to the introduction of the concept
of scaling and self-affinity\ref{9}. It has been verified
experimentally\ref{10} that the height variations $z$ of rock and
metal surfaces are invariant within a certain range under a
scale transformation $z\rightarrow\lambda^\zeta z$ where
the roughness exponent $\zeta$ is quite universally around
0.85. This scale invariance is the origin of anomalous elastic
behaviour\ref{11}.

Geometrically the scale invariance means that one has asperities
on all length scales. In particular one can decompose the
topography of the surface into different scales of resolution
and then expect that the behaviour be the same on each scale.
We will use this idea in our approach to write down a closed
expression for the friction as a function of the driving
velocity for a simple model surface.
Our principal assumption will be that the friction parameter $\mu $ is
the same on all levels. This is a kind of mean field assumption
which does not take into account the scaling exponent $\zeta$.
In fact, one can also think of formulating a renormalization procedure
that describes how to go from one scale to the next using $\zeta$
and this work is in progress\ref{12}.

In our model we make several simplifying assumptions concerning the
geometry of the surface. First we reduce the problem from three
to two dimensions as if the surfaces were made out of grooves
so that considering a cross section would be enough. Furthermore
we suppose that
all the grooves on one scale have the same size and
spacing, and that they are identical for both lids.
A schematic plot is shown in fig.~1. We consider
asperities that just have
a triangular shape of an inclination angle $\alpha$ and width $B$.

Gravity $g$ pushes the two bodies together while they are
sheared against each other with a fixed relative velocity $v$.
We suppose that at the areas where the two surfaces are in contact
the Coulomb law holds so that the force
acting against the direction of the relative motion
is proportional to the normal force with proportionality constant $\mu$.
Therefore this friction force has a magnitude
$$f_{fr} = m \mu g \ cos(\alpha) \pm F \mu sin(\alpha)\eqno(2)$$
where the second term is due to the component of the force $F$
pulling the two surfaces against each other.
The force of eq.~(2) is independent on the contact area
which therefore drops out entirely of the problem and as a
consequence the size of the considered system becomes irrelevant
in our calculation.

When the bodies now move on top of each
other they will jump up and down. When they move up their surfaces slide
against each other as shown in fig.~1 (phase I of the motion)
until the tips of the triangles exactly touch.
After that they will fly down on a
parabolic trajectory (phase II of the motion)
until the surfaces hit again against each other.
They can hit either on the left side (case $a$)
or on the right side (case $b$) of the triangles on the lower surface.
In the calculation all the three cases
(I, II$a$ and II$b$) must be considered
separately. Moreover one has to take into account that the upper
lid must not necessarily hit the lower one within the next
period $B$ but can jump over $N (N \ge 1)$
periods (fig.~2). Only in phase I a restoring force acts which can be
calculated to be
$$
F(\mu ) =  m \cdot g \cdot {sin(\alpha ) + \mu \cdot
cos(\alpha ) \over cos(\alpha ) - \mu \cdot sin(\alpha )} \eqno(3)
$$
In order to know how much time this force acts one must calculate
the point $x_s$ at which the surfaces hit against
each other. These are in case $a$
$$
x_s = {3 \over 2} \cdot B + v^2 \cdot \varphi + v \cdot \sqrt{\varphi ^2 \cdot
v^2 - \varphi \cdot (N-1) \cdot B} \eqno(4a)
$$
and in case $b$
$$
x_s = {3 \over 2} \cdot B + v \cdot \sqrt{B \cdot \varphi \cdot N} \eqno(4b)
$$
with
$$
\varphi = {2 \cdot tan(\alpha ) \over g}
$$
According to our previous discussion we now assume that also
the entire system fulfills the Coulomb relation (at fixed
velocity) with the same friction constant $\mu$ because
in our picture, when looking with a microscope, the flat surfaces
of the triangles are themselves made out of many small triangles
following the principles of self-similarity.
This gives us the self-consistency equation
$$
f = {1 \over T} \cdot \int_0^T F(\mu ,t ) dt = m \mu g\eqno(5)
$$
where $T = N \cdot B/v$ is the period, i.e. the time to go once through
phases I and II. Solving eq.~5 we get
$$
\mu = {1 \over 2 \cdot tan(\alpha )} \Bigl[{1
\over N} ({1\over 2} + \kappa )-1 \Bigr] + \sqrt{{1 \over 4 \cdot
tan^2(\alpha )} \Bigl[{1
\over N} ({1\over 2} + \kappa )-1 \Bigr]^2 + {1
\over N} \Bigl[ {1 \over 2}-\kappa \Bigr]}\eqno(6a)
$$
with
$$\kappa = N - {1 \over 2} - {v
\over B} \cdot (\sqrt{\varphi ^2 \cdot v^2 - \varphi \cdot (N-1)
\cdot B} +\varphi \cdot v)
$$
if
$$
v \in \Biggl[ \sqrt{{B \over \varphi}} \cdot \sqrt{N-1} \ , \
\sqrt{B \over \varphi} \cdot {N-{1 \over 2} \over
\sqrt{N}} \Biggr]
$$
(case $a$) and
$$\mu = {\lambda \over \sqrt{tan(\alpha)}} +
\sqrt{{\lambda ^2 \over tan( \alpha )} + \lambda \cdot 2 \cdot \sqrt{
tan(\alpha )}-1}\eqno(6b)$$
with
$$ \lambda = v \cdot \sqrt{{1 \over 2 \cdot B \cdot g \cdot N}}$$
otherwise (case $b$).

In fig.~3 we show the $\mu$ of eq.~(6)
as a function of $v$ for various values
of $\alpha$. We clearly see the two cases $a$ and $b$:
When the surfaces hit on the left (right) side of the triangle
$\mu$ increases (decreases) with $v$. In between there is the
case at which one has jumps from tip to tip and in that case
$\mu$ vanishes. This happens at the characteristic velocity
of the system which for dimensional reasons is $\sqrt{Bg}$.
When the angle increases the friction
increases and the $a$-$b$ sequences become shorter (the jumps become
longer). For small velocities and small angles Coulomb's
original proposition seems fulfilled: $\mu$ does not depend on $v$.
However, for very small velocities $\mu$ falls off to zero. This
effect which gets stronger with $\alpha$ is an artifact of our
self-consistent approach and therefore only rather small
angles $\alpha$ (less that $5^{\circ}$) are allowed.

The zig zag behaviour of the curves in fig.~3 is due to the
commensurability of the indentations of the two surfaces.
In reality surfaces have random asperities and no synchronized
motion as in our model is possible. In order to take care of this
heterogeneity we now average the values of $\mu$ over an entire
range of angles $\alpha$ changing $B$ such that the height of the
triangles remains the same. This can be seen in
fig.~4 for the case in which we average over all $\alpha$
between $0.2^{\circ}$ and $2.0^{\circ}$ with the same
probability for each angle. As expected one finds a smooth
behaviour. In fig.~4 we have fitted these data to a curve of the form
$\mu = (v+v_0)^\beta$ with $v_0=3.308$ and $\beta = -2.342$
and find very good agreement.
We also tried several other possible functional forms for $\mu$
including the ones of eq.~1. In Fig.~5 we see how well they can be
described by our model by showing the difference between our
values and the best fits to a large range of velocities.
An Ansatz due to eq.~1a is fitted using
$v_0=-1.198 $ and $\beta = 0.0224$ and gives reasonable agreement
while the analogous fit using
eq.~1b shows deviations beyond the scale of fig.~5 and
only works well in a smaller range of $v$.
Other functional forms like ${\rm log}^{\beta}((v+v_0)$
and $v_0 \cdot v^{\beta}$ in fact fit very well too.

We have presented a very simple geometrical model in order to
calculate the friction coefficient from
the steric hindrances of the contacting surfaces.
By assuming the same friction coefficient on different scales
we can formulate a self-consistent equation. We take disorder into
account by a very crude averaging procedure. The results of this
approximate calculation are in qualitative
agreement with the observed behaviour but clearly also other
functional forms for the velocity dependence of $\mu$ work
very well. In fact, we might expect that the various  expressions
would work equally well as assumptions for earthquake models.

Our purely geometrical approach has an analogy to
the case of granular media in which traditionally a phenomenological
static friction between the grains is introduced\ref{13}
in order to describe the finite angle of repose of a
heap\ref{14}. Recently, however,
the insight was gained that the angles of repose can also
be controlled just by varying the
shape of the grains\ref{15}
in a model with no static friction. Therefore also
for granular media the friction can be described only by
the geometry of the asperities (grain shapes).
\bigskip
\bigskip
\leftline{\bf Acknowledgement}
\smallskip
We thank the HLRZ team for a very enlightening discussion
at the Kaffeeseminar and S. Roux and J.P. Vilotte for useful comments.
\vfill\eject
\noindent{\bf References}
\medskip
\item{1.} C. de Coulomb, {\it M\'emoires de Math\'ematiques
et de Physique} (L'Imprimerie Royale, Paris, 1773), p.343
\item{2.} K.L. Johnson, {\it Contact Mechanics} (Cambridge
Univ. Press, 1989)
\item{3.} D.F. Moore, {\it Principles and Applications
of Tribology} (Pergamon, Oxford, 1975)
\item{4.} J.M. Carlson and J.S. Langer, Phys. Rev. Lett.
{\bf 62} (1989), 2632 and Phys. Rev. A {\bf 40} (1989), 6470
\item{5.} R. Burridge and L. Knopoff, Bull. Seismol. Soc.
{\bf 57} (1967), 341
\item{6.} J.H. Dietrich, J. Geophys. Res. {\bf 84} (1979), 2161
and 2169; A.L. Ruina, J. Geophys. Res.
{\bf 88} (1983), 10359
\item{7.} J.R. Rice and A.L. Ruina, Trans. ASME, J. Appl. Mech.
{\bf 50} (1983), 343; J.R. Rice, preprint
\item{8.} H.J. Herrmann, G. Mantica and D. Bessis, Phys. Rev. Lett.
{\bf 65} (1990), 3223
\item{9.} F. Tzschichholz and M. Pfuff, in {\it Fracture Processes in
Concrete, Rock and Ceramics}, eds. J.G.M. van Mier, J.G. Rots and
A. Bakker, Vol~1 (Cambridge Univ. Press., 1991)
\item{10.} E. Bouchaud, G. Lapasset and J. Plan\'es, Europhys. Lett.
{\bf 13} (1991), 73
\item{11.} S. Roux, J. Schmittbuhl,
J. P. Vilotte and A. Hansen, preprint;
K.J. Mal\o y, X.-l. Wu, A. Hansen and S. Roux, preprint
\item{12.} C. Moukarzel, T. P\"oschel and H.J. Herrmann, in progress
\item{13.} P. Cundall and O.D.L. Strack, G\' eotechnique {\bf 29}
(1979), 47
\item{14.} J. Lee and H.J. Herrmann, J. Phys. A {\bf 26} (1993), 373
\item{15.} J.A.C. Gallas and S. Soko\l owski, Int. J. Mod. Phys. B
{\bf 7} (1993), 2037;
V. Buchholtz and T. P\"oschel, preprint HLRZ 33/93
\vfill\eject
\noindent{\bf Figure Captions}
\medskip
\item{\bf Fig.~1} Schematic plot of our model.
\item{\bf Fig.~2} Phases of the motion: in phases ($a$) and ($b$)
the trajectories end on the right and left sides of the triangles
respectively after jumping over $N$ triangles.
\item{\bf Fig.~3} Friction coefficient $\mu$ as a function of
the driving velocity $v$ for different angles $\alpha$
and $B = 0.0035 / tan (\alpha)$. We always use $g = 9.81$.
\item{\bf Fig.~4} Friction coefficient $\mu$ as a function
of $v$ homogeneously
averaged over $\alpha$-values that range from
$0.2^{\circ}$ to $2.0^{\circ}$.
Superposed we see the function
$\mu = (v+v_0)^\beta$ with $v_0=3.308$ and $\beta = -2.342$
fitted for $2.2 \le v \le 70.$
\item{\bf Fig.~5} Difference between our result for $\mu(v)$ and the
best fits to various possible functional
forms ${\cal F}(v)$ plotted as function
of $v$, corresponding to ${\cal F}(v) = {\rm log}^{\beta}((v+v_0)$
with $v_0 = 4.431$ and $\beta = -6.183$;
${\cal F} = (v+v_0)^\beta$ with $v_0=3.308$ and $\beta = -2.342$;
${\cal F} = \beta/(v+v_0)$ with $v_0 = -1.198$ and $\beta = 0.0224$ and
${\cal F} = v_0 \cdot v^{\beta}$ with $v_0 = 0.0645$ and
$\beta = -1.438$, fitted for $2.2 \le v \le 70.$
\end

{}.